\documentclass{moriond}

\def\Journal#1#2#3#4{{#1} {\bf #2}, #3 (#4)}

\def\NIMA{{\em Nucl. Instrum. Methods} A}

\def\JPG{{\em J. Phys.} G}

\def\PRL{\em Phys. Rev. Lett.}
\def\PRD{{\em Phys. Rev.} D}

\def\JHEP{\em JHEP}
\def\PTEP{\em Prog. Theor. Exp. Phys.}

\def\be{\begin{equation}}
\def\ee{\end{equation}}
\def\bea{\begin{eqnarray*}}
\def\eea{\end{eqnarray*}}

\newcommand{\Photo}{\includegraphics[height=35mm]{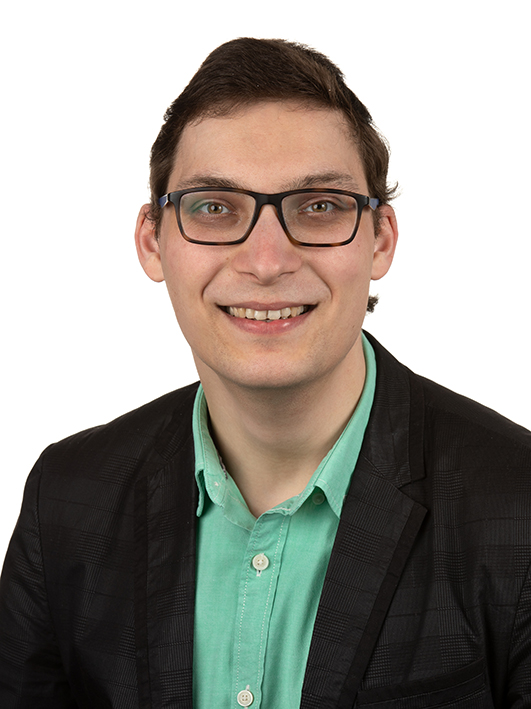}}

\begin{document}
\vspace*{4cm}
\title{Rare decays and searches for exotic signatures at LHCb: purely leptonic decays}

\author{ T. Mombächer on behalf of the LHCb Collaboration}

\address{Instituto Galego de Física de Altas Enerxías, Universidade de Santiago de Compostela,\\
Rúa de Xoaquín Díaz de Rábago s/n,\\
15782 Santiago de Compostela, Galicia, Spain}

\maketitle\abstracts{
	A review of the experimental status of purely leptonic \mbox{${B^0_{[s]}\to\ell^+\ell^-}$} decays is given and the new analysis of \mbox{${B^0_{[s]}\to\mu^+\mu^-}$} observables with the LHCb experiment is presented. The new results are the most precise single experiment measurements of the quantities to date and are consistent with the Standard Model. However, they also match the recent anomalies found in \mbox{$b\to s\ell\ell$} transitions excellently.
}

\section{Introduction}
Rare decays provide a very clean way of probing the Standard Model (SM). Because the SM background is small, potential processes beyond the Standard Model (BSM) can have a large impact on the rates and properties of the decays. 
In recent years, measurements of decays via \mbox{$b\to s\ell\ell$} transitions have shown tantalising deviations from the SM predictions.
These decays are suppressed in the SM because they can only proceed via flavour changing neutral currents. 
While the deviations from the SM are only moderately significant individually, they form a consistent pattern when considered with the tools of effective field theory (see e.g. efforts of global fits~\cite{stangl}).
They favour a modification of vector and axial-vector couplings with respect to the SM.
In the context of effective field theory fits the purely leptonic \mbox{${B^0_{[s]}\to\ell^+\ell^-}$} decays play a special role, since they are sensitive to only axial-vector couplings in the SM.
Compared to the other decays of the \mbox{$b\to s\ell\ell$} kind they are additionally helicity suppressed, making them even rarer in the SM.
In the presence of BSM processes, also right-handed couplings, as well as pseudoscalar and scalar couplings can contribute to the decays, which would lift the helicity suppression.

\section{State of the art in \boldmath${B^0_{[s]}\to\ell^+\ell^-}$ decays}
\begin{sloppypar}
Because of its purely leptonic final state and owing to strong theory efforts, very accurate SM predictions of the \mbox{$B^0_{[s]}\to\ell^+\ell^-$} branching fractions of~\cite{Bobeth,Fleischer} \mbox{$\mathcal{B}(B^0_{[s]}\to\tau^+\tau^-)=(2.14\pm0.12)\times10^{-8}$} \mbox{$[(7.56\pm0.35)\times10^{-7}]$}, \mbox{$\mathcal{B}(B^0_{[s]}\to\mu^+\mu^-)=(1.03\pm0.05)\times10^{-10}$} \mbox{$[(3.66\pm0.14)\times10^{-9}]$}, and \mbox{$\mathcal{B}(B^0_{[s]}\to e^+e^-)=(2.41\pm0.13)\times10^{-15}$} \mbox{$[(8.60\pm0.36)\times10^{-14}]$} can be achieved.
From the experimental point of view the branching fractions are much less precisely known.
The \mbox{${B^0_{[s]}\to \tau^+\tau^-}$} and \mbox{${B^0_{[s]}\to e^+e^-}$} decays have not been observed yet. The most sensitive upper limits have been determined with the LHCb experiment to the values \mbox{$\mathcal{B}(B^0_{[s]}\to\tau^+\tau^-)<2.1\times10^{-3}\ [6.8\times10^{-3}]$} and \mbox{$\mathcal{B}(B^0_{[s]}\to e^+e^-)<3.0\times10^{-9}\ [11.2\times10^{-9}]$} at \mbox{$95\,\%$} CL~\cite{tau,e}. 
Although the experimental upper limits are far away from the SM predictions, they are still probing the parameter space of BSM models. While the tauonic modes are especially sensitive to models that include violation of lepton flavour universality, the electronic decay modes can strongly probe effects from models introducing scalar and pseudoscalar currents, where the strong helicity suppression would be lifted.
The \mbox{${B^0_{[s]}\to\mu^+\mu^-}$} decays, however, are experimentally accessible. The mode \mbox{${B^0_{s}\to\mu^+\mu^-}$} has been unambiguously observed with several experiments and its branching fraction determination is becoming a precision measurement. However, the mode \mbox{${B^0\to\mu^+\mu^-}$} is still unconfirmed experimentally.
A recent combination of analyses by the ATLAS, LHCb and CMS collaborations~\cite{bmm-comb} yielded \mbox{$\mathcal{B}(B^0_{s}\to\mu^+\mu^-)=2.69^{+0.37}_{-0.35}\times10^{-9}$} and \mbox{$\mathcal{B}(B^0\to \mu^+\mu^-)<1.9\times10^{-10}$} at \mbox{$95\,\%$} CL. 
Although the sizeable uncertainties make the result still inconclusive, it is interesting to note that the experimental \mbox{$\mathcal{B}(B^0_{s}\to\mu^+\mu^-)$} value is lower than the SM prediction in the same way as found in other branching fraction measurements of \mbox{$b\to s\mu^+\mu^-$} decays~\cite{iso,L,K*,phi} and lepton flavour universality ratios~\cite{RK*,RpK,RK}.
With the clear observation of \mbox{${B^0_{s}\to\mu^+\mu^-}$} decays also their behaviour concerning charge-parity ($CP$) transformations can be investigated, which can be strongly modified by BSM processes even if their rate is SM-like.
In the SM, only the heavy ($CP$-odd) \mbox{$B^0_{s}$} mass eigenstate can decay into two muons. Because of the sizeable lifetime difference in the \mbox{$B^0_s$} system between \mbox{$1.620\,\rm{ps}$} (heavy eigenstate) and \mbox{$1.423\,\rm{ps}$} (light eigenstate, $CP$-even)~\cite{pdg}, a measurement of the effective lifetime of the \mbox{${B^0_{s}\to\mu^+\mu^-}$} decay exhibits sensitivity to distinguish the two \mbox{$B^0_{s}$} eigenstates.
Experimentally the value is determined to \mbox{$\tau_{B^0_{s}\to\mu^+\mu^-}=1.91^{+0.37}_{-0.35}\,\rm{ps}$} excluding the measurement presented in these proceedings from a combination of measurements with the CMS and LHCb experiments~\cite{bmm-comb} and is compatible with both eigenstates.
\end{sloppypar}

\section{The new analysis of \boldmath$B^0_{[s]}\to\mu^+\mu^-$ decays with the LHCb experiment}
\begin{figure}
\centerline{\includegraphics[width=0.45\linewidth]{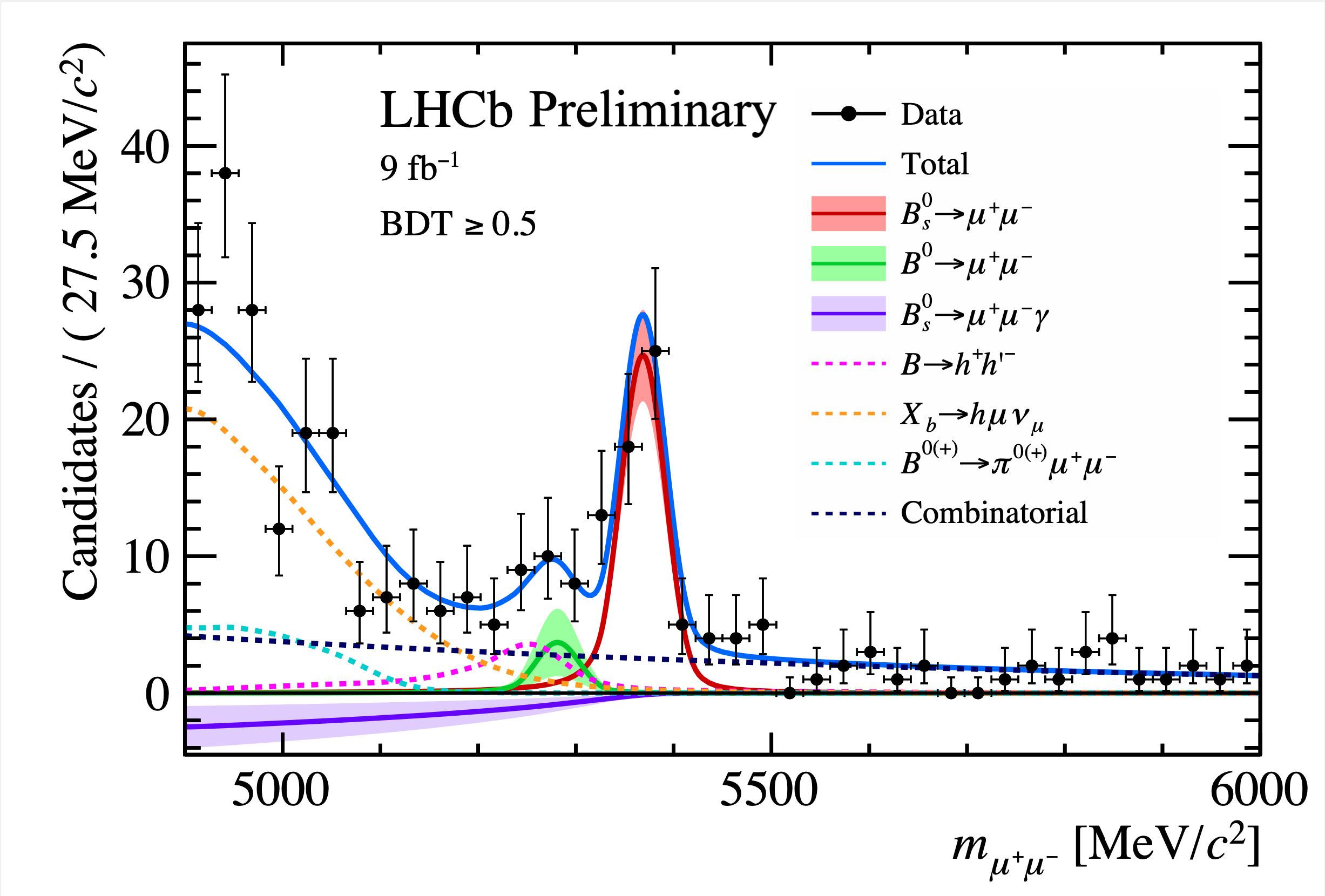}}
\caption[]{Mass distribution of the selected dimuon candidates (black dots) with \mbox{$\rm{BDT}>0.5$}. The result of the fit is overlaid and the different components are detailed: \mbox{$B^0_{s}\to\mu^+\mu^-$} (red solid line), \mbox{$B^0\to\mu^+\mu^-$} (green solid line), \mbox{$B^0_{s}\to\mu^+\mu^-\gamma$} (violet solid line), combinatorial background (blue dashed line), \mbox{$B\to h^+h^{\prime-}$} (magenta dashed line), \mbox{$B^0\to\pi^-\mu^+\nu_\mu$}, \mbox{$B^0_s\to K^-\mu^+\nu_\mu$}, \mbox{$B^+_c\to J\!/\!\psi\mu^+\nu_\mu$} and \mbox{$\Lambda^0_b\to p\mu^-\bar{\nu}_\mu$} (orange dashed line), and \mbox{$B^{0(+)}\to\pi^{0(+)}\mu^+\mu^-$} (cyan dashed line). The solid bands around the signal shapes represent the variation of the branching fractions by their total uncertainty.}
\label{fig:fitresult}
\end{figure}

A new analysis of \mbox{${B^0_{[s]}\to\mu^+\mu^-}$} observables has been performed with the full LHCb data set collected in the LHC Runs 1 and 2 between 2011 and 2018~\cite{bmm-prl,bmm-prd}.
This corresponds to a total luminosity of about \mbox{$9\,\rm{fb}^{-1}$} of proton-proton collisions at center-of-mass energies of \mbox{$7,8,\rm{\ and\ } 13\,\rm{TeV}$}, effectively doubling the size of the data set investigated in a previous analysis~\cite{bmm-old}.
The new analysis does not only determine the branching fractions of \mbox{${B^0_{[s]}\to\mu^+\mu^-}$} decays and the effective lifetime \mbox{$\tau_{B^0_{s}\to\mu^+\mu^-}$}, but also for the first time searches for initial state radiation \mbox{${B^0_{s}\to\mu^+\mu^-\gamma}$} decays with \mbox{$m_{\mu^+\mu^-}>4.9\,\rm{GeV}/c^2$} without reconstructing the photon. This decay is expected to have a branching fraction of \mbox{$\mathcal{O}(10^{-10})$} in the SM~\cite{Kozachuk}.
The overall selection strategy is kept similar to the previous analysis. Following a fully inclusive trigger selection, a very efficient preselection on kinematical and topological variables is used to reduce the data set to a manageable size. Strong particle identification requirements suppress exclusive background decays with misidentified particles and backgrounds from \mbox{$B^+_c\to J\!/\!\psi(\mu^+\mu^-)\mu^+\nu_\mu$} decays are specifically vetoed via the invariant mass of combining the signal muon with any other muon in the event.
The final separation between signal and different background components is achieved by fitting the invariant dimuon mass distribution in bins of a high performant boosted decision tree (BDT), where the BDT is trained on topological variables, most notably the isolation of muons with respect to other tracks in the event.
While the selection strategy remains unchanged with respect to the previous analysis, large improvements are achieved in the calibration of the signal simulation, as well as the determination of the misidentification probability of pions and kaons as muons in data.
The signal BDT distributions are taken from simulation, which has been corrected for \mbox{$B$} kinematics and detector occupancy variables using data control modes. Additional corrections for trigger and particle identification mismodellings obtained from data are also applied. 
This method is compared to the approach used in the previous analysis, the direct calibration of the BDT distribution from \mbox{$B\to h^+h^{\prime-}$} decays in data and found to be in excellent agreement yet significantly improved precision.
The signal mass peak is determined from \mbox{$B\to h^+h^{\prime-}$} decays in data as well, while the mass resolution is interploated from various quarkonia resonances.
The particle identification and misidentification probabilities are determined from high statistics calibration modes in data.
Most important here is the determination of the hadron-muon misidentification probability, as that determines the size of backgrounds from \mbox{$B\to h^+h^{\prime-}$} decays, which are peaking in the signal region.
Therefore the hadron-muon misidentification probability is independently checked against \mbox{$B\to h^+h^{\prime-}$} decays in data, where one of the hadrons is identified as a muon.
The \mbox{${B^0_{[s]}\to\mu^+\mu^-(\gamma)}$} branching fractions are determined relative to two normalisation modes, \mbox{$B^0\to K^+\pi^-$} and \mbox{$B^+\to J\!/\!\psi(\mu^+\mu^-) K^+$}. Since the normalisation modes are \mbox{$B^+$} and \mbox{$B^0$} decays, for the signal \mbox{$B^0_s$} decays the fragmentation fraction ratio of \mbox{$B^0_s$} and \mbox{$B^0$} mesons, \mbox{$f_s/f_d$} has to be considered, which received an improvement in precision by a factor of 2 in a recent combination of several LHCb measurements~\cite{fsfd}.
The signal components are distinguished against several background components in the fit to the invariant mass, which are modelled individually: combinatorial background is described by a falling exponential distribution; \mbox{$B\to h^+h^{\prime-}$} decays with two misidentified hadrons, \mbox{$B^0\to\pi^-\mu^+\nu_\mu$}, \mbox{$B^0_s\to K^-\mu^+\nu_\mu$} and  \mbox{$\Lambda^0_b\to p\mu^-\bar{\nu}_\mu$} decays with one misidentified hadron and partially reconstructed \mbox{$B^{0(+)}\to\pi^{0(+)}\mu^+\mu^-$} and \mbox{$B^+_c\to J\!/\!\psi(\mu^+\mu^-)\mu^+\nu_\mu$} decays are modelled from simulation with Gaussian kernel distributions.
The fit result is displayed in Fig.~\ref{fig:fitresult}, exhibiting a clear \mbox{$B^0_{s}\to\mu^+\mu^-$} signal with a significance of about \mbox{$11\sigma$}.
The \mbox{$B^0\to\mu^+\mu^-$} and \mbox{$B^0_{s}\to\mu^+\mu^-\gamma$} signals are found to be compatible with zero at \mbox{$1.7\sigma$} and \mbox{$1.5\sigma$}, respectively. Therefore upper limits are calculated with the CL$_s$ method~\cite{cls}.
The branching fractions measured are \mbox{$\mathcal{B}({B^0_{s}\to\mu^+\mu^-})=\left(3.09^{\,+\,0.46\,+\,0.15}_{\,-\,0.43\,-\,0.11}\right)\times 10^{-9}$}, \mbox{$\mathcal{B}({B^0\to\mu^+\mu^-})<2.6\times 10^{-10}$}, and \mbox{$\mathcal{B}({B^0_{s}\to\mu^+\mu^-\gamma;m_{\mu^+\mu^-}>4.9\,\rm{GeV}/c^2})<2.0\times 10^{-9}$} at \mbox{$95\,\%$} CL.
The \mbox{$B^0_{s}\to\mu^+\mu^-$} branching fraction is slightly lower but consistent with the SM prediction.
While the measurement is statistically dominated, its most relevant systematic uncertainties come from the normalisation branching fractions and \mbox{$f_s/f_d$}, both having a contribution of about \mbox{$3\,\%$}.

The effective lifetime of \mbox{$B^0_{s}\to\mu^+\mu^-$} decays is measured with a similar selection strategy as for the branching fraction measurement with some modifications:
in order to enhance the signal efficiency, the particle identification requirements are softened; the BDT distribution is binned into only two regions in order to simplify the analysis; the dimuon invariant mass region is restricted to be above \mbox{$5320\,\rm{MeV}/c^2$}, which allows to neglect all exclusive backgrounds but the combinatorial background.
With a fit to the dimuon invariant mass backgrounds are statistically subtracted using the {\em sPlot} method~\cite{splot}. 
The background-subtracted decay time distribution is then fitted simultaneously in two BDT regions to obtain the effective lifetime, where the acceptance introduced by the selection is modelled from simulation.
The procedure is tested with \mbox{$B^0_s\to K^+K^-$} and \mbox{$B^0\to K^+\pi^-$} decays, finding good agreement to measurements previously obtained with the LHCb experiment.
From the simultaneous fit an effective lifetime of \mbox{$\tau_{B^0_{s}\to\mu^+\mu^-}=2.07\pm0.29\pm0.03\,\rm{ps}$} is found, which is consistent with the heavy \mbox{$B^0_s$} eigenstate (SM hypothesis) at \mbox{$1.5\sigma$} and the light eigenstate at \mbox{$2.2\sigma$}.
The precision of this measurement improves over the previous LHC combination and is fully statistically dominated, where systematic uncertainties mainly arise due to the limited knowledge of the decay time acceptance shape. 

\section{Summary}
\begin{sloppypar}
In the context of \mbox{$b\to s\ell\ell$} transitions purely leptonic \mbox{${B^0_{[s]}\to\ell^+\ell^-}$} decays play a special role, since in the SM they are sensitive to axial-vector couplings only. Particularly the measurements of \mbox{$B^0_{[s]}\to\mu^+\mu^-$} observables are becoming precision measurements.
The newly presented \mbox{$B^0_{[s]}\to\mu^+\mu^-$} analysis with the LHCb experiment provides the most precise single experiment measurement of the branching fraction and effective lifetime of \mbox{$B^0_{s}\to\mu^+\mu^-$} decays.
Furthermore tight upper limits are imposed on the branching fraction of \mbox{$B^0\to\mu^+\mu^-$} decays and a first ever search for initial state radiation \mbox{$B^0_{s}\to\mu^+\mu^-\gamma$} decays at high dimuon invariant mass has been performed.
The \mbox{$B^0_{[s]}\to\mu^+\mu^-$} branching fraction is slightly lower, yet compatible with the SM. However, it is fully in agreement with the picture emerging from the anomalies in other \mbox{$b\to s\ell\ell$} decays. 
The LHCb detector is currently being upgraded and is expected to resume data taking with an increased rate, which will soon allow to establish, whether the picture seen in other \mbox{$b\to s\ell\ell$} decays is confirmed in \mbox{$B^0_{[s]}\to\mu^+\mu^-$} decays.
\end{sloppypar}


\end{document}